\documentclass{article}

\usepackage{arxiv}

\usepackage[utf8]{inputenc} 
\usepackage[T1]{fontenc}    
\usepackage{hyperref}       
\usepackage{url}            
\usepackage{booktabs}       
\usepackage{amsfonts}       
\usepackage{nicefrac}       
\usepackage{microtype}      
\usepackage{lipsum}		
\usepackage{graphicx}
\usepackage{natbib}
\usepackage{doi}
\usepackage{threeparttable} 
\usepackage{amsmath,bm} 
\usepackage{bbm} 
\usepackage{subcaption} 
\usepackage{derivative} 
\usepackage{xcolor}

\title{Heteroscedasticity-aware stratified sampling to improve uplift modeling}

\author{ \href{https://orcid.org/0000-0000-0000-0000}{\includegraphics[scale=0.06]{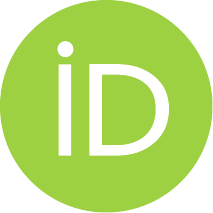}\hspace{1mm}Björn Bokelmann} \\
	Chair of information systems\\
	Humboldt University Berlin\\
	Unter den Linden 6\\
    10099 Berlin\\
	\texttt{bokelmab@hu-berlin.de} \\
	\And
	\href{https://orcid.org/0000-0000-0000-0000}{\includegraphics[scale=0.06]{orcid.pdf}\hspace{1mm}Stefan Lessmann} \\
	Chair of information systems\\
	Humboldt University Berlin\\
	Unter den Linden 6\\
    10099 Berlin\\
}

\renewcommand{\shorttitle}{Heteroscedasticity in uplift modeling}

\hypersetup{
pdftitle={Binarized target modeling},
pdfsubject={q-bio.NC, q-bio.QM},
pdfauthor={Björn~Bokelmann, Stefan~Lessmann},
pdfkeywords={First keyword, Second keyword, More},
}

\begin{document}
\maketitle

\begin{abstract}
In many business applications, including online marketing and customer churn prevention, randomized controlled trials (RCT's) are conducted to investigate on the effect of specific treatment (coupon offers, advertisement mailings,...). Such RCT's allow for the estimation of average treatment effects as well as the training of (uplift) models for the heterogeneity of treatment effects between individuals. The problem with these RCT's is that they are costly and this cost increases with the number of individuals included into the RCT. For this reason, there is research how to conduct experiments involving a small number of individuals while still obtaining precise treatment effect estimates. We contribute to this literature a \textit{heteroskedasticity-aware stratified sampling} (HS) scheme, which leverages the fact that different individuals have different noise levels in their outcome and precise treatment effect estimation requires more observations from the "high-noise" individuals than from the "low-noise" individuals. By theory as well as by empirical experiments, we demonstrate that our HS-sampling yields significantly more precise estimates of the ATE, improves uplift models and makes their evaluation more reliable compared to RCT data sampled completely randomly. Due to the relative ease of application and the significant benefits, we expect HS-sampling to be valuable in many real-world applications.       
\end{abstract}


\section{Introduction}
The estimation of treatment effects is of highest importance in a wide range of applications: Medical institutions need to infer the effect of new treatments on the patient populations \citep{foster2011subgroup}, online shops need to assess the effect of marketing incentives like advertisement and coupons on the purchase behavior of their customers\citep{haupt2022targeting} and companies with contract customers need to assess the effect of anti-retention measures on the probability of their customers to churn\citep{verbeke2012new}.

Treatment effects on an outcome of interest (like survival, purchase amount or churn status) are the change in this outcome due to the treatment, compared to the case that no treatment was provided. In most applications, two kinds of treatment effects are of particular interest: The average treatment effect (\textit{ATE}) is an average of the treatment effects of each individual in a population. Knowledge of the average treatment effect is helpful, because it helps to judge, whether the treatment has a relevant beneficial effect or not. The second kind of treatment effect is the conditional average treatment effect (\textit{CATE}). In contrast to the ATE, the CATE takes into account that the treatment effect will be different between different individuals. The CATE measures the treatment effect conditional on individual's features (like age, gender, previous online purchase behavior or health issues). The estimation of CATE requires statistical model building. This model building and the the application of the model to achieve a most beneficial treatment allocation over the individuals is referred to as \textit{uplift modeling}. Thereby "uplift" is named the effect of the treatment on an individual customer, which can "lift up" the outcome compared to the case that this customer was not treated.

Estimation of treatment effects (be it ATE or CATE) is difficult due to the fundamental problem of causal inference, which is that for each individual one can either measure the outcome if this individual received the treatment or the outcome if this individual did not receive the treatment. The most reliable method to infer treatment effects is to perform a randomized controlled trial (\textit{RCT}). In an RCT, a sample of the whole population of customers/patients is chosen for the experiment. Within this sample, treatment is randomly assigned to some individuals, while the other individuals do not receive the treatment. The group of individuals with treatment is called \textit{treatment group} and the group of individuals without treatment is called \textit{control group}. Such RCT's are frequently conducted by companies and there is vast literature about the estimation of ATE or CATE from such RCT's.\citep{larsen2023statistical,jin2023toward,arbour2021efficient} The problem with RCT's is the cost due to sub-optimal treatment assignment decisions: Individuals without notable effect have the same chance to receive treatment than individuals with high beneficial treatment effect.\citep{arbour2021efficient} This cost grows with the sample size of the RCT (number of individuals included) and accordingly, there is a strong incentive to keep such experiments as small as possible, while still obtaining reliable effect estimates.

There are various approaches to reduce the sample size of RCT's in previous literature: Covariate adjustment \citep{deng2013improving} and stratification \citep{xie2016improving} aim at an ATE estimation, with reduced variance in the ATE estimate. This reduced variance offers the opportunity to perform RCT's with smaller sample size, while still obtaining meaningful results. Covariate balancing techniques aim for a more similar features distribution in the treatment and the control group than would be the case without.\citep{kallus2018optimal} This higher similarity of feature distributions also reduces the variance in ATE and CATE estimates and can thereby help to reduce the sample size of RCT's. Another approach is active learning, where the idea is to iteratively collect the samples which are most helpful for the effect estimators.\citep{sundin2019active} Again this approach allows to conduct experiments of smaller sample size while providing reliable effect estimates, compared to a random selection of a sample for the RCT. 

In this research, we also aim at a reduction of the RCT's sample size, by systematically choosing the right sample. Our idea is to use the concept of \textit{heteroskedasticity}: Different individuals in the population will have a different noise level in their outcomes. To get reliable effect estimates, it is necessary to have a high number of observations of individuals with high noise level, while for indivduals with low noise level a smaller number of observations is sufficient. Following this principle, we collect RCT data, where individuals with a higher noise level appear with higher proportion than in the whole population, while individuals with a low noise level appear with a lower proportion. As we demonstrate in this paper, this kind of \textit{heteroskedasticity-aware stratified sampling (HS sampling)} leads to more precise ATE and CATE estimates (and thereby the opportunity to reduce the RCT sample size, while maintaining reliable results). Our research differs from the previous literature in three aspects: First, we aim at a reduction of another source of variance, namely the noise on individual level, than other variance reduction techniques. For this reason our approach reliably leads to additional variance reduction, when it is combined with other methods like covariate adjustment. Second, we aim at a complete selection of individuals for the RCT, before the experiment has started. This poses an application advantage compared to sample selection methods like active learning, which iteratively select individuals during the experiment. Third, we consider the important aspect of CATE model evaluation and show that our approach provides significant improvements there. This aspect is import, because estimation of CATE is only useful, when the precision of according models can be assessed. 

Our approach is applicable under the following three conditions, which are arguably often met in business-related applications of RCT's: (1) The outcome of interest is a binary variable. (2) The outcome rate in the whole population of customers is low (far lower than 50\%). (3) There is pre-experimental data of untreated individuals, based on which one can train a model for the outcomes of costumers. Regarding the condition (1), it is easy to see that most outcomes applied in RCT's of previous literature are binary (e.g. purchase yes/no, churn yes/no, page view yes/no), so our approach remains relevant in most applications. Regarding condition (2) it is also easy to see, that this condition is mostly met: purchase rates are mostly low and so are churn rates and page view rates. Condition (3) might pose the highest application challenge, but arguably it could be also fulfilled for a wide range of applications: In churn prevention, it will mostly be possible to model some heterogeneity in churn probability between customers, based on information like age, or length of subscription. In the same way, online shops are also likely to have outcome models for their customer base. Hence, we expect our HS-sampling to be widely applicable.


\section{Related literature}

\subsection{Uplift modeling and CATE estimation}
The \textit{uplift modeling} literature provides methods to support decision which individuals to treat, in order to maximize (economic) benefits. Such applications include, among others, the allocations of coupons or other incentives in online marketing \citep{gubela2020response,verbeke2023or,haupt2022targeting} and the allocation of anti-churn measures to individuals \citep{devriendt2021you}. In such applications, the goal shifts from ATE estimation, to CATE estimation, where heterogeneity in the treatment effects between individuals is taken into account, by estimating treatment effects conditional on the individuals features. Note, that the application-oriented uplift modeling literature branch is not the only literature branch concerned with CATE estimation. There is also the \textit{heterogeneous treatment effect estimation} literature branch, which is less application-oriented and routed in the literature about statistics and econometrics.\citep{athey2019generalized,chernozhukov2018double,nie2021quasi} Accordingly, most of our considerations also apply for this branch of literature. However, the heterogeneous treatment effect estimation literature is not much focused on CATE estimation on RCT data. Since our ideas only concern sample selction for RCT's, our research is more relevant to the uplift modeling branch. In this literature branch, it is common to train and evaluate uplift models on RCT data.\citep{haupt2019affordable} The big problem is, that, since uplift modeling (or equivalently CATE estimation) and its evaluation is a notoriously unstable process, there is a demand for large RCT's to generate reliable models and obtain meaningful model evaluation results.\citep{fernandez2022causal,bokelmann2023improving} 

Since this demand of large RCT's is a big obstacle in the application of uplift models (due to costs and logistical problems) there is research about how to conduct experiments with smaller sample sizes, while still obtaining useful uplift models.  A solution to this problem is active learning for CATE estimation. Those studies start with a given training data set, which could be RCT data \citep{connolly2023task} or observational data \citep{sundin2019active,qin2021budgeted} and provide methods to iteratively include further observations in the training data, which are expected to be most beneficial for the CATE estimators. This targeted selection of samples enables to achieve the same precision in CATE estimates by using a smaller experimental sample than for an RCT with random choice of samples. Most of these studies obtain new samples in a sequential manner \citep{jesson2021causal}, but some studies also consider batch-wise active learning \citep{puha2020batch}. The idea of active learning is related to our research in that we also seek for a way to obtain precise uplift models, based on a small data set, by selective sampling. However, in contrast to the active learning studies, our sample selection entirely takes place before the experiment has started and any treatment is provided. In addition, our study differs from these active learning studies in that we not only seek for precision of uplift models but also seek for reliable evaluation of their performance.   



\subsection{Stratification and other pre-experimental sampling techniques}\label{sec_rel_lit_strat}
In contrast to active learning, pre-experimental sampling techniques choose samples before the experiment has started. Most of the pre-experimental sampling techniques aim at a reduction of the variance of ATE estimates by restricting the randomness in the sampling process. One well-established method of variance reduction is stratified sampling. The idea is to pre-define some strata (groups of individuals defined by certain feature characteristics). Then individuals are included in the RCT, according to certain pre-defined proportions. In most cases, these sample proportions are chosen as the proportions of the strata in the whole population.\citep{xie2016improving}, \citep{deng2013improving},\citep{berman2022latent},\citep{barrios2014optimal},\citep{,aufenanger2017machine}. There is research about the optimization of stratified sampling. \citet{bai2022optimality} derives optimal strata, if pre-experiment data is available. \citet{tabord2023stratification} suggests a two-stage method where optimal strata a learned during a first stage of an experiment and applied to the second stage. In the context of survey statistics, \citet{cytrynbaum2023optimal} derives an optimal solution for the definition of strata, the proportion of individuals sampled from these strata and the treatment proportion within these strata. Our HS-sampling approach is strongly related to the idea of stratified sampling. However, apart from proposing a customized approach to find suitable stratification parameter our research differs from these previous studies in that we use stratified estimation not only to reduce the ATE estimator variance but also to improve uplift modeling. 

Another well-established RCT sampling procedure is \textit{covariate (feature) balancing}. Covariate-balancing techniques aim to achieve a balanced distribution of features among treatment and control group. The studies by  \citet{greevy2004optimal,harshaw2023balancing,imai2008variance,kallus2018optimal,addanki2022sample} suggest different approaches towards covariate balancing, where the basic idea is always to restrict the randomness of the treatment allocation to some extend, such that a certain degree of similarity in the feature distribution between treatment and control group is guaranteed. Importantly, this does not change the fundamental property of an RCT, namely that the probability of an individual to obtain treatment is equal for all individuals in the sample.

While all the former studies suggested pre-experiment sampling approaches to improve ATE estimation, there are also a few studies suggesting sampling approaches to improve uplift modeling. \citet{haupt2019affordable} provide a method of semi-random treatment assignment, where the probability of treatment is increased for individuals beeing expected to have a high positive treatment effect. This approach is related to our idea, in that a pre-experimental model is used to guide the design of the experiment. However, in contrast to \citet{haupt2019affordable}, we apply a pre-experimental outcome model (and no uplift model) and seek at an RCT sample with most beneficial statistical properties for uplift modeling instead of a sample, where the treatment allocation decision was made under economical considerations. Another sampling approach for improving uplift modeling is provided by \citet{arbour2022online}. The authors calculate leverage scores for all individuals in the customer base and decide, based on these scores, which individuals to include in an RCT to most efficiently build uplift models. Our approach differs from this idea in that we use pre-experimental outcome information and choose the sample according to heteroskedasticity considerations. 


\subsection{Post-experiment variance reduction techniques}
Once the RCT was conducted and the data was gathered, there is the incentive to estimated the ATE with as little variance (or equivalently as much precision) as possible. Previous literature provides methods for such post-experiment variance reduction techniques for ATE estimation. There are two common approaches: \textit{stratified estimation} and \textit{covariate adjustment}. Stratified estimation is always applied to estimate the ATE, when stratified sampling was conducted. Hence, in all the studies applying stratified sampling in section \ref{sec_rel_lit_strat}, stratified estimation is also applied as a means for variance reduction. The second common variance reduction method is covariate adjustment. Covariate adjustment aims to adjust differences between treatment and control group in the conditional expected values, which are not due to the treatment but due to sampling. A traditional method for covariate adjustment in online experiments is CUPED, where a linear regression model is used to adjust for such differences in the conditional expected values between treatment and control group.\citep{deng2013improving} More recently, machine learning methods for covariate adjustment have been proposed.\citep{guo2021machine,hosseini2019unbiased,cohen2020no,jin2023toward} Thereby \citet{jin2023toward} suggest a covariate adjustment procedure, which asymptotically leads to the optimal variance reduction, as long as the applied machine learning algorithms are consistent. The idea of covariate adjustment is not only applicable to ATE estimation, but it is also possible to apply it for variance reduction in the uplift model evaluation.\citep{bokelmann2023improving} 

A third, less common approach, which we found in the literature about online RCT's is variance weighted ATE estimation, where the noise level in the individual's outcomes is estimated based on pre-experiment data. The observations used for the ATE estimate are then inversely weighted by this estimated noise level, which results in variance reduction, but bares the risk of bias.\citep{liou2020variance} This approach is related to our idea, in that heterogeneity in the noise level (heteroskedasticity) is used to make more efficient effect estimates. Our approach differs from this weighted ATE estimation in that we also suggest a modified sampling scheme and aim for improving uplift model training and evaluation in addition to ATE estimation. 

\section{Estimation of treatment effects}
In this research, we consider RCT data, which consists for each individual of a binary outcome $y\in\{0,1\}$, the features $x$, and the treatment status $w\in\{0,1\}$ (where $w=1$ denotes treatment). Per definition of an RCT, the probability to receive a treatment is independent of the features. On such a data set, the following relationship holds
\begin{align}
    y=\mu_{x}+w\cdot\tau_x+\varepsilon, \label{eq_rep_outcome}
\end{align} where $\mu_x:=E[y|w=0,x]$ is the conditional probability of a positive outcome for an untreated individual and
\begin{align*}
    \tau_x:=E[y|w=1,x]-E[y|w=0,x]
\end{align*} is the conditional average treatment effect (CATE). The noise $\varepsilon$ denotes the variation in $y$, which can not be explained by the features $x$ and the treatment status $w$. 

As our goal is the estimation of treatment effects, anything except for $\tau_x$, which affects the outcome, makes the task statistically harder. Notably, the components $\mu_x$ and $\varepsilon$ can be seen as a nuisance. Statistically, their effect on treatment effect estimation procedures is a variance increase. In consequence, these estimation procedures become unreliable if $\mu_x$ and $\varepsilon$ contribute strongly to the variance of $y$ and the sample size of the RCT is low.    

\subsection{Average treatment effect (ATE) estimation}
The average treatment effect (ATE) is defined by
\begin{align*}
    ATE=E[y|w=1]-E[y|w=0].
\end{align*} 

The important property of RCT data is that the \textit{difference-in-means} estimator
\begin{align*}
    \hat{ATE}=\frac{1}{N_w}\sum_{w_i=1}y_i-\frac{1}{N_{\bar{w}}}\sum_{w_i=0}y_i,
\end{align*} where $N_w,N_{\bar{w}}$ denote the number of treated respectively untreated individuals in the sample, is unbiased for the ATE. In this way, medical institutions can estimate treatment effects of medication on patients and companies can estimate treatment effects of e.g. providing coupons or showing advertisement on customers.

When performing an RCT to estimate the ATE, one needs to be sure that the resulting estimate $\hat{ATE}$ is precise. How precise the estimate $\hat{ATE}$ is expected to be is determined by its variance
\begin{align*}
    Var[\hat{ATE}]=\frac{1}{N}\left(\frac{Var[y|w=1]}{p}+\frac{Var[y|w=0]}{(1-p)}\right).
\end{align*} Here $p$ is the proportion of treated individuals in the sample, such that $N_w=p\cdot N$ and $N_{\bar{w}}=(1-p)\cdot N$. $N=N_w+N_{\bar{w}}$ denotes the number of all individuals in the sample. As we can see, the variance $Var[\hat{ATE}]$ decreases with increasing sample size $N$.    

The problem with such RCT's is that their cost usually increases with the sample size $N$. This cost is either because providing the treatment is costly or holding back treatment on individuals reduces potential gains. The total cost increases accordingly with the number of individuals $N$ included in the RCT sample. Hence, there is an incentive to perform RCT's with a small sample size $N$, while still having a low variance in the ATE estimate. This is why there is vast research about \textit{variance reduction methods} for RCT's. Such statistical methods can reduce the variance of a treatment effect estimator, without increasing the RCT sample size. The left plot of figure \ref{fig:treat_var} illustrates the problem of variance for the ATE estimation and the potential use of methods for variance reduction. In the next section, we describe the variance reduction procedure "stratified sampling and estimation", which is most relevant for our research. In Appendix \ref{app_variance_red_techniques} we provide a detailed description of other variance reduction techniques.

\subsection{Stratified sampling and estimation}\label{sec_strat_samp}
The idea of stratified sampling is to divide the feature space in exclusive strata (groups). For simplicity, we show the principle of stratification for two strata $S_H,S_L$. To estimate the average treatment effect, one needs to know the proportion $p_H$ of individuals with features in stratum $S_H$ in the whole population. The estimator then applied is
\begin{align}
    \hat{ATE}_{S}=p_{H}\cdot\hat{\tau}_{H}+(1-p_H)\cdot\hat{\tau}_{L}\label{eq_ate_s},
\end{align} where $\hat{\tau}_{H},\hat{\tau}_{L}$ denote the difference-in-means estimators in stratum $S_H$ respectively $S_L$. When choosing a cohort for the experiment, one needs to decide the proportion $\frac{N_H}{N}$ of individuals from stratum $S_H$ sampled in the whole sample of size $N$. The most common approach is \textit{proportional sampling}, where one chooses $\frac{N_H}{N}=p_H$ as the proportion of individuals in $S_H$ in the whole population. This stratified sampling and estimation procedure leads to a variance reduction compared to the random sampling-based estimator $\hat{ATE}$, if the conditional expected value of the outcome is different between both strata. For details, we refer to Appendix \ref{app_strat}.


In most previous applications of stratified sampling and estimation in RCT's, proportional sampling is applied \citep{xie2016improving,berman2022latent,aufenanger2017machine,barrios2014optimal}. However, it is also possible to choose $\frac{N_H}{N}$ differently, while still maintaining an unbiased estimator using equation \eqref{eq_ate_s}. An alternative to proportional sampling would be \textit{optimal allocation} (aka Neyman allocation). To explain the idea of optimal allocation, it is usefull to introduce the notation $V_i=\frac{Var[y|w=1,S_i]}{p}+\frac{Var[y|w=0,S_i]}{1-p}$ with $i=H,L$, for the outcome variance within the strata. If one of the strata has a higher outcome variance than the other stratum, the variance of the estimator in equation \eqref{eq_ate_s} can be reduced, if individuals from the high variance stratum (for simplicity we say $S_H$) get sampled unproportionally often. We define the \textit{oversampling ratio} of the high variance stratum as $R_H:=(N_H/N)/p_H$. Optimal allocation would apply the oversampling ratio, which minimizes the variance of the estimator in equation \eqref{eq_ate_s}. We found this kind of sampling scheme in recent literature about survey statistics.\citep{cytrynbaum2023optimal} The optimal oversampling ratio would be
\begin{align}
    R_H=\left(p_H+\frac{1-p_H}{\sqrt{Q_{V}}}\right)^{-1},\label{eq_opt_proportion}
\end{align} where $Q_{V}:=V_H/V_L$ denotes the variance quotient. If this oversampling ratio is chosen, the quotient of the variance for the ATE estimator with optimal allocation and the variance for the ATE estimator with proportional sampling would be  
\begin{align}
    \frac{Var[\hat{ATE}_{OS}]}{Var[\hat{ATE}_{S}]}=\frac{\left(p_H\cdot \sqrt{Q_{V}}+(1-p_H)\right)^2}{p_H\cdot Q_{V}+(1-p_H)}\label{eq_var_sopt}.
\end{align} As $Q_V>1$, there is always a variance reduction due to optimal allocation.

\subsection{Uplift modeling and CATE estimation}
Uplift modeling (CATE estimation) is a more fine-grained task than ATE estimation. The CATE $\tau_x$ from equation \eqref{eq_rep_outcome} is a function of the features $x$. Hence, statistical model building is required to estimate it. A wide variety of methods exists, to train an uplift model $\hat{\tau}(x)$ on RCT data. In this research, we only analyse three well established methods, namely the two-model approach (T-learner) \citep{hansotia2002incremental}, the single-model approach (S-learner) \citep{hill2011bayesian,foster2011subgroup} and the X-learner\citep{kunzel2019metalearners}. These methods are called "meta-learners", because they are based on conventional supervised learning models and use these models to obtain predictions for $\tau_x$. For details, we refer to the provided references.

The T-learner estimates the CATE by taking the difference of two supervised learning model's predictions 
\begin{align*}
    \hat{\tau}_{T}(x):=\hat{\mu}_{1}(x)-\hat{\mu}_{0}(x).
\end{align*} Thereby, $\hat{\mu}_{1}(x)$ denotes an outcome model trained on the treated individuals and  $\hat{\mu}_{0}(x)$ denotes an outcome model trained on the untreated individuals. The S-learner estimates the CATE by only using one outcome model according to
\begin{align*}
    \hat{\tau}_{S}(x):=\hat{\mu}(x,w=1)-\hat{\mu}(x,w=0).
\end{align*} In contrast to the T-learner, the underlying outcome model $\hat{\mu}(x,w)$ is trained on treated as well as untreated individuals and has the treatment status $w$ as a feature. The X-learner is a bit more complicated than the former two approaches. In a first step, it requires transformation of the outcomes of the treated individuals in the data set to $y^{1}:=y-\hat{\mu}_0(x)$ and transformation of the outcomes of untreated individuals to $y^{0}:=\hat{\mu}_1(x)-y$. In the second step, it requires to train one supervised learning model $\hat{\tau}^{1}_{X}(x)$ on the treated individuals with transformed outcomes and one supervised learning model $\hat{\tau}^{0}_{X}(x)$ on the untreated individuals with transformed outcomes. The CATE estimates are than generated from these two models by
\begin{align*}
    \hat{\tau}_{X}(x)=p\cdot \hat{\tau}^{1}_{X}(x)+(1-p)\cdot\hat{\tau}^{0}_{X}(x). 
\end{align*}


\subsection{Uplift model evaluation}\label{sec_cate_eval}
As CATE estimates are always model-based, some caution is required. Due to the possibility of bias or over-fitting, model-based predictions for the CATE can be far of the real CATE values. This is a difference to ATE estimation, where the precision of estimates follows from simple statistical considerations. As a consequence, uplift models should always be evaluated before applying them to support treatment decisions or draw any conclusion about treatment effect heterogeneity in the population. 

There are different principles how to evaluate uplift models. One way is to asses mean squared deviations $E[(\hat{\tau}(x)-\tau_x)^2]$ between the CATE estimates and the actual CATE values. This principle is mostly applied in the heterogenous treatment effect estimation literature branch. According metrics provide information about the precision of CATE estimates, but do not measure the economic impact of uplift model based treatment decisions. Another way to evaluate uplift models is to measure how well they can rank individuals by their treatment effect. \citet{radcliffe2007using} suggested the Qini curve, which measures for each $t\in [0,1]$ the cumulative treatment effect one would obtain when providing treatment to the share $t$ highest ranked individuals. This evaluation principle measures the economic impact of using an uplift model for treatment decisions and is mostly applied in the uplift modeling literature branch. As we mainly follow the uplift modeling perspective on CATE estimation, we choose the Qini curve as our evaluation metric of interest.    

As described by \citet{bokelmann2023improving}, the Qini curve suggested by \citet{radcliffe2007using} corresponds to an estimator of
\begin{align*}
     ATE_{t}\cdot N_w,
\end{align*} where $ATE_{t}$ denotes the ATE on the share $t\in [0,1]$ of highest ranked individuals by an uplift model and $N_w$ is the number of treated individuals within this share of highest ranked individuals. In the Qini curve definition provided by \citet{radcliffe2007using}, the scale of the curve would depend on the sample size $N$ of the test data (because $N_w$ depends for each $t\in [0,1]$ on this size $N$). So, it can be useful to remove this sample size depends by dividing by $p\cdot N$. The modified Qini curve version is then
\begin{align}
    Q(t)=ATE_{t}\cdot t, \label{eq_Qini}
\end{align} where we use that $\frac{N_w}{p\cdot N}\approx t$ on RCT data.

The Qini curve is a good way to visually assess the CATE model performance. For comparing the performance of different models, which we do in our computational experiment, it is usefull to have a numerical measure for the performance. A typically applied measure is the \textit{area under the Qini curve} (AUQ).\citep{devriendt2020learning} In Appendix \ref{app_cate_evaluation}, we describe in detail how the AUQ can be calculated based on experimental and real-world data. 

\begin{figure}
\centering
\centering
  \includegraphics[width=.9\linewidth]{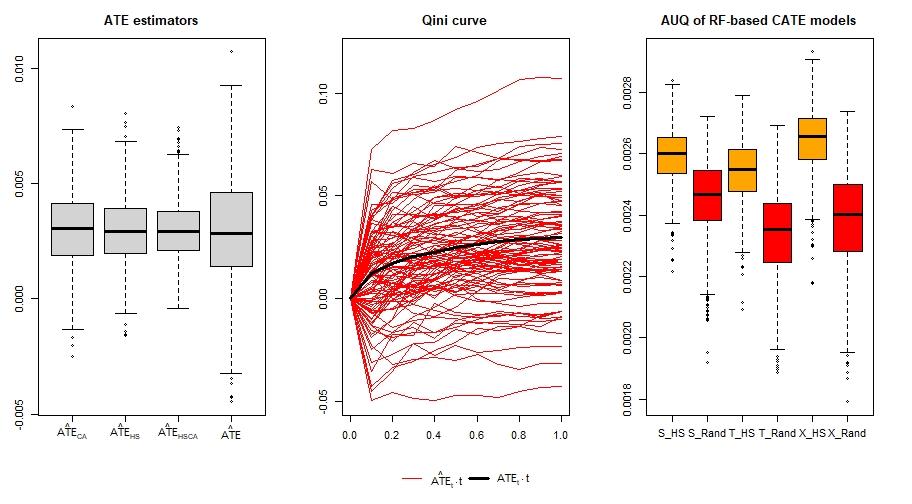}
  \captionof{figure}{The problem of variance in treatment effect estimation. The plots illustrate empirical results of our simulation scenario 3. The left figure illustrates by box-plots the variance of different ATE estimation procedures between 1,000 repetitions of the experiment. $\hat{ATE}_{CA},\hat{ATE}_{HS},\hat{ATE}_{CA}$ denote the ATE estimator based on covariate adjustment, HS-sampling and the combinition of HS-sampling and covariate adjustment respectively. $\hat{ATE}$ denotes the difference-in-means estimate on completely randomly sampled RCT data. The middle plot illustrates the variance in the estimation of the Qini curve between 100 reptitions of the experiment. The right plot illustrates by box-plot the variance in the performance of uplift modeling methods (T-,S-,X-learner, each trained on HS-respectively completely randomly sampled data) between 1,000 repetitions of the experiment.}
  \label{fig:treat_var}
\end{figure}

We close this section by noting that the CATE model evaluation by the Qini curve is essentially an ATE estimation task. Hence, any ideas about the problem of variance and the methods of variance reduction are expected to the evaluation of CATE models. For illustration purpose, we refer to the middle plot in figure \ref{fig:treat_var}, which demonstrate the problem of variance when applying the Qini curve for uplift model evaluation.

\section{Heteroskedasticity-aware stratified sampling (HS sampling)}\label{sec_hs_sampling}
Our \textit{heteroskedasticity-aware stratified sampling} (abriviated HS sampling in the following) is based on the principle of stratified sampling with optimal allocation, described in section \ref{sec_strat_samp}. Our idea involves the definition of two strata $S_H,S_L$, where $S_H$ includes individuals with a high expected outcome value and $S_L$ includes individuals with a low expected outcome values. The name "heteroskedasticity aware" comes from the following fact: The noise level of an untreated individual with binary outcome is given by 
\begin{align*}
    Var[\varepsilon|x,w=0]=E[y|x,w=0]\cdot (1-E[y|x,w=0]).
\end{align*} This noise level is the higher, the closer $E[y|xw=0]$ is to 50\%. As we consider application settings, with low $E[y,w=0]$, we can be relatively certain that most individuals have $E[y|x,w=0]<0.5$ and so we can assume that the noise level tends to grow with $E[y|x,w=0]$.  

Our stratification should tend to put individuals with high $E[y|x,w=0]$ in $S_H$ and with low $E[y|x,w=0]$ in $S_L$. Hence, we would expect the noise level in $S_H$ to be much higher than in $S_L$. This is the reason for the name "heteroskedasticity aware". Sampling individuals from $S_H$ with a high proportion is usefull, because these individuals have a high noise level and so many observations are required for statistically reliable treatment effect estimation procedures. 

Note, that the noise level of an individual also depends on the treatment status, because $E[y|x,w=1]\neq E[y|x,w=0]$ and, up until now, we have only discussed the case of untreated individuals. In our approach, we make the simplifying assumption $Var[y|x,w=1]\approx Var[y|x,w=0]$. There are two reasons for this assumption: First, the strata definition and sampling process takes place before the experiment. So, no information about the treatment effect is available at this point and the best proxy for the noise level of a treated individual is the expected noise level of the same individual having received no treatment. Second, we expect the effect of the treatment on the noise level to be rather small in most practical applications. Treatment might change the conditional expected value of individuals slightly but unless the heterogeneity in $\tau_x$ is much stronger than the heterogeneity in $\mu_x$ our assumption about variance similarity is remains justified. In any way, problems about our assumption would become evident in the empirical evaluation of the HS-sampling approach.    

The whole HS-sampling procedure is illustrated in figure \ref{fig:hs_sampling_illustration}. This procedure involves 7 steps and is described in the following subsection.  

\subsection{HS-sampling procedure}\label{sec_choice_ratio}
\begin{figure}
\centering
\centering
  \includegraphics[width=.9\linewidth]{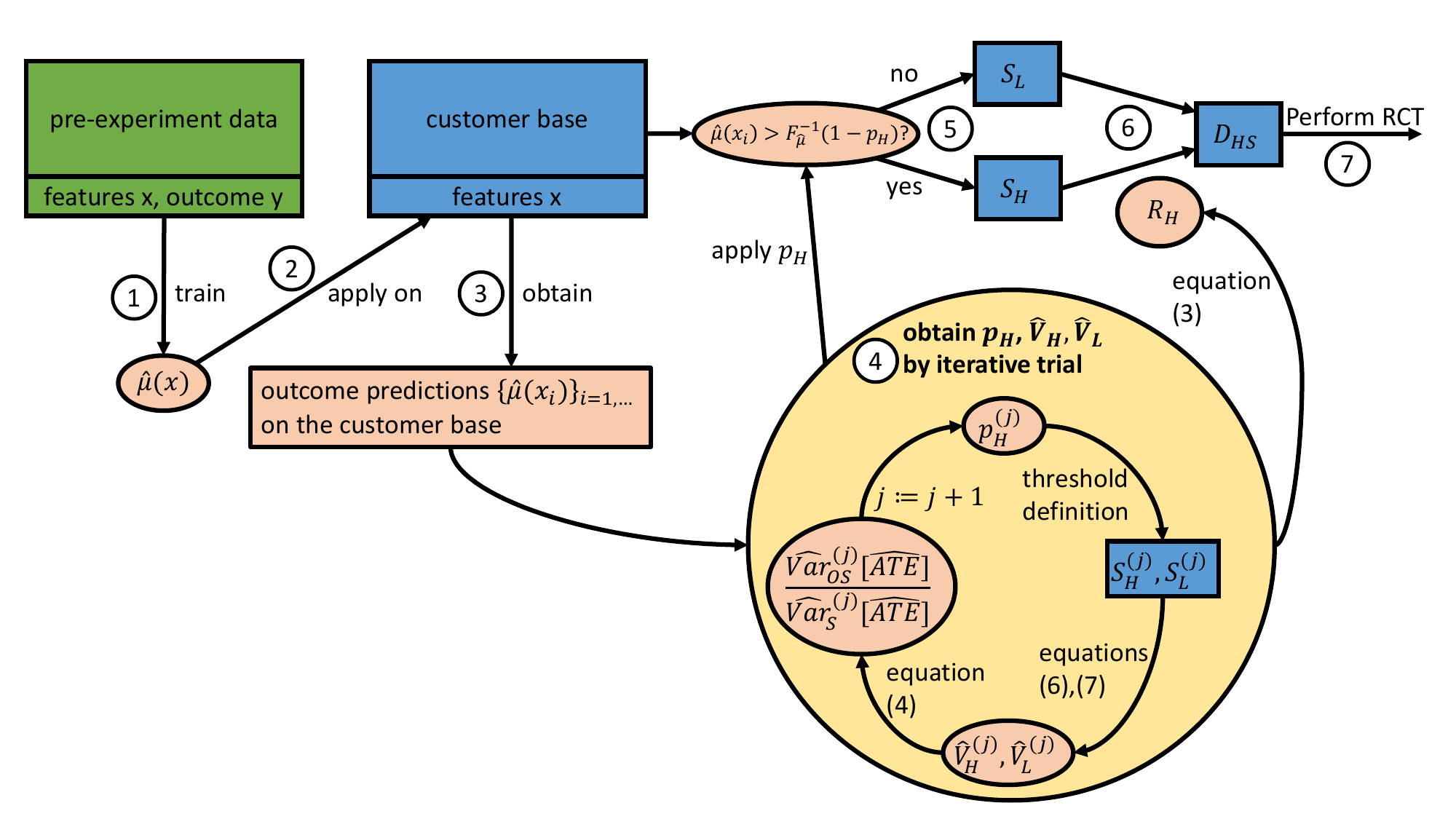}
  \captionof{figure}{Illustration of the HS-sampling procedure. The description of the procedure is given in section \ref{sec_choice_ratio}.}
  \label{fig:hs_sampling_illustration}
\end{figure}

Step (1) of our HS-sampling procedure involves the training of an outcome model $\hat{\mu}(x)$ for the expected probability of a positive outcome $\mu_x$ in the control group (see figure \ref{fig:hs_sampling_illustration}). To train such a model, pre-experimental data of untreated customers involving features $x$ and outcome values $y$ is required.

In steps (2) and (3) of the procedure, the outcome model $\hat{\mu}(x)$ is applied on the customer base (from which individuals for the RCT could be sampled) to obtain for the $i=1,...$ individuals in the customer base predictions $\{\hat{\mu}(x_i),\}_{i=1,...}$. For this step, it is necessary that the features used to train $\hat{\mu}(x)$ on the pre-experimental data are also available for the individuals in the customer base.

Step (4) needs some more explanation. It involves an iterative search procedure for the optimal definition of strata $S_H,S_L$ as well as the optimal oversampling ratio $R_H$ of individuals from the high variance stratum. Regarding the definition of strata $S_H,S_L$, we use the predictions $\{\hat{\mu}(x_i),\}_{i=1,...}$ to rank the individuals from low to high expected outcome values. Having ranked the customer base in this way, we only need to decide about a threshold such that individuals are placed in $S_H$ respectively $S_L$, depending on whether their predictions exceed this threshold. For each $p_H\in [0,1]$, the quantile $F^{-1}_{\hat{\mu}}(1-p_H)$ would place individuals in strata $S_H$ and $S_L$ with proportions $p_H$ respectively $(1-p_H)$. 

To find the best definition of $S_H$ and $S_L$, we thus iteratively try 99 values $p^{(j)}_H=\frac{j}{100}$ for $j\in\{1,2,...,99\}$. Each value $p^{(j)}_H$ yields a distinct definition of the strata $S^{(j)}_H$ and $S^{(j)}_L$. Their respective outcome variances are given by 
\begin{align}
    V^{(j)}_{H}&=E\left[\mu_x\middle|\hat{\mu}(x)>F^{-1}_{\hat{\mu}}(1-p^{(j)}_H)\right]\cdot \left(1-E\left[\mu_x\middle|\hat{\mu}(x)>F^{-1}_{\hat{\mu}}(1-p^{(j)}_H)\right]\right)\label{eq_VH}\\
    V^{(j)}_{L}&=E\left[\mu_x\middle|\hat{\mu}(x)\leq F^{-1}_{\hat{\mu}}(1-p^{(j)}_H)\right]\cdot \left(1-E\left[\mu_x\middle|\hat{\mu}(x)\leq F^{-1}_{\hat{\mu}}(1-p^{(j)}_H)\right]\right).\label{eq_VL}
\end{align} These variance values are not observable, so we need estimates $\hat{V}^{(j)}_H,\hat{V}^{(j)}_L$ for them. To get such estimates, we simply replace the conditional expected value $\mu_x$ in the above equations by our model estimates $\hat{\mu}(x)$ and estimate the according expected values by sample averages. Now, with our proportion value $p^{(j)}_H$ and stratum-dependent outcome variance estimates $\hat{V}^{(j)}_H,\hat{V}^{(j)}_L$, we can build the quotient $\hat{Q}^{(j)}_{V}:=\frac{\hat{V}^{(j)}_H}{\hat{V}^{(j)}_L}$ and then estimate the variance reduction if optimal allocation is applied with equation \eqref{eq_var_sopt}. By following this procedure for each $p^{(j)}_H$ with $j\in\{1,2,...,99\}$, we can plot the estimated variance reduction for each $p^{(j)}_H$. An illustration of such a plot is provided in figure \ref{fig:hs_potential}. From this plot, we obtain the proportion $p_H$, which we expect to yield the optimal stratification. With this value of $p_H$, we obtain the threshold $F^{-1}_{\hat{\mu}}(1-p_H)$ for our final strata definition. To get an oversampling ratio $R_H$ for stratum $S_H$, we plug $p_H$ and the corresponding variance quotient estimate  
$\hat{Q}_V=\frac{\hat{V}_H}{\hat{V}_L}$ into formula \eqref{eq_opt_proportion}. This finalizes step (4) and yields all parameters necessary to determine the HS-sampling scheme.

Steps (5) and (6) involve dividing the customer base into the strata $S_H$,$S_L$ and sampling from these strata with proportions $R_H\cdot p_H$ and $1-R_H\cdot p_H$ respectively. The final step (7) then simply involves random treatment allocation, such that an RCT is performed. 

\subsection{Robustness and practical considerations}\label{sec_hs_robust}
\begin{figure}
\centering
\centering
  \includegraphics[width=.9\linewidth]{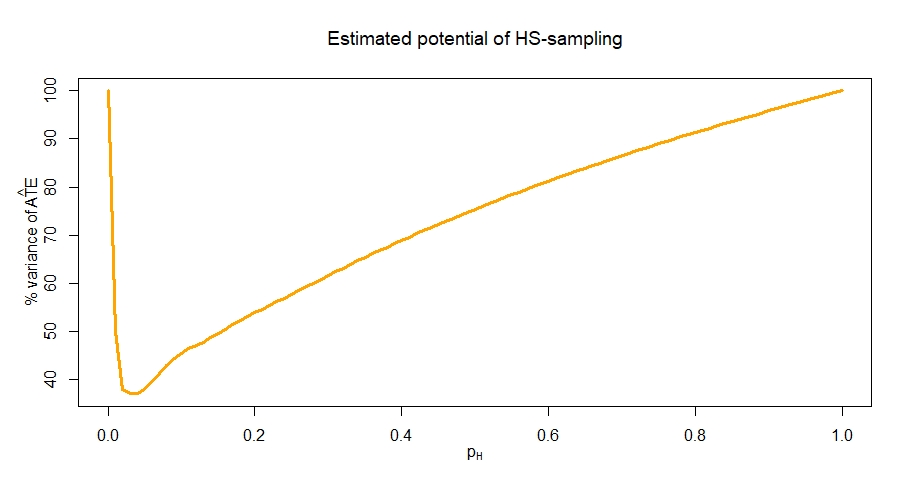}
  \captionof{figure}{The estimated variance of an HS-sampling based ATE estimator in simulation scenario 3. The estimated variance is calculated using equation \eqref{eq_var_sopt}, with the $\hat{\mu}(x)$-based strata variance estimates $\hat{V}_{H},\hat{V}_{L}$ plugged-in for various values of $p_H\in [0,1]$. Based on these result, we would expect up to 63\% variance reduction by HS-sampling. 
  }
  \label{fig:hs_potential}
\end{figure}

HS-sampling relies on one fundamental assumption, namely that we can build an outcome model $\hat{\mu}(x)$, based on pre-experimental data and can apply this model to our customer base. Clearly, this requires the that the pre-experimental features used for the model building are also available for the model application on the customer base. However, this is not a very strong requirement. One only needs to build $\hat{\mu}(x)$ on the intersection of features available in the pre-experimental data and the customer base data. Having trained such an outcome model, one can asses the potential of HS-sampling in terms of variance reduction, by applying step (4) of the HS-procedure. An illustration of a resulting plot for the potential benefits is given in figure \ref{fig:hs_potential}. 

\begin{figure}
\centering
\centering
  \includegraphics[width=.9\linewidth]{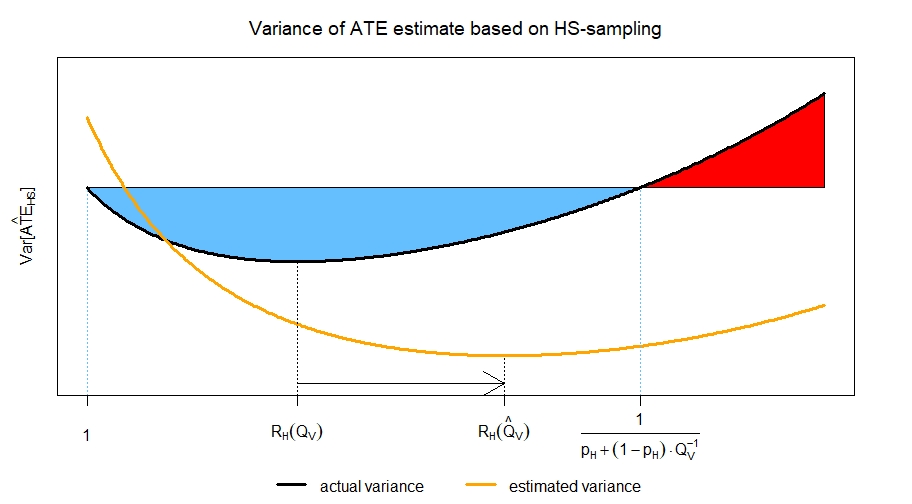}
  \captionof{figure}{The effect of the oversampling ratio $R_H$ on the ATE estimator variance. The estimated optimal oversampling ratio $R_H(\hat{Q}_V)$ is higher than the actual optimum $R_H(Q_V)$, due to prediction errors. As long as the chosen oversampling ratio is within $\left[1,\frac{1}{p_H+(1-p_H)\cdot Q_V^{-1}}\right]$, there is a variance reduction due to HS-sampling. For higher values of $R_H$, there is a variance increase due to HS-sampling. 
  }
  \label{fig:oversampling}
\end{figure}

If a significant potential for variance reduction is recognized, the question becomes: What can potentially go wrong when applying HS-sampling? To answer this question, it is helpful to use our illustration in figure \ref{fig:oversampling}. For a chosen $p_H$, the oversampling ratio $R_H$ needs to be determined. If we choose $R_H=1$, the sampling scheme corresponds to conventional stratified sampling with proportional allocation and no harm is done. The variance reduction effect of HS-sampling sets in once $R_H$ surpasses 1. Using equation \eqref{eq_var_strat_unprop}, we can derive that HS-sampling leads to a variance reduction, if 
\begin{align*}
    R_H\in\left[1,\frac{1}{p_H+(1-p_H)\cdot Q_V^{-1}}\right].
\end{align*} In our HS-sampling procedure, we choose $R_H$ based on the estimated variance quotient $\hat{Q}_V$, which is generated by the outcome model $\hat{\mu}(x)$. The only thing which can go wrong with HS-sampling is that $R_H(\hat{Q}_V)$ is chosen too high, such that a variance increase is caused. In figure \ref{fig:oversampling}, this is marked by the red area.

In Appendix \ref{app_robustness}, we analyze potential problems due to prediction errors of $\hat{\mu}(x)$ in detail. We argue why one can expect the chosen oversampling ratio $R_H(\hat{Q}_V)$ to be higher than the optimal oversampling ratio $R_H(Q_V)$. As a countermeasure, we suggest to use adjusted values
\begin{align*}
    p_H^{ad}&:=p_H+\frac{1}{4}p_H\\
    R_H^{ad}&:=\frac{3}{4}R_H(\hat{Q}_V)+\frac{1}{4}\cdot 1.
\end{align*} We apply these adjusted values in our computational experiment, where we demonstrate the application of HS-sampling. In Appendix \ref{app_robustness}, we also perform a simulation study which shows that the problem of choosing $R_H(\hat{Q}_V)$ too high happens if $\hat{\mu}(x)$ is chosen with high complexity (and has hence a tendency to over-fit). The simulation study also shows that our proposed adjustment works very well to prevent a harmful application of HS-sampling.  

\subsection{Uplift model training and evaluation}\label{sec_hs_cate_est}
We have seen that for the estimation of the ATE, a weighted estimation procedure is necessary (see equation \eqref{eq_ate_s}), if HS-sampling is applied. This is because due to the oversampling from $S_H$ the feature distribution in the experiment differs from the feature distribution in the whole population. Interestingly, CATE estimation does not require a modification of the estimation procedure. This is because the CATE $\tau_x$ is a treatment effect conditional on the feature value $x$. If some feature values are more frequently sampled in the experimental set, this only yields a higher uplift model precision at these points but does not cause a bias. Hence, uplift model training can be done in the same way as studied in previous literature.

In contrast, uplift model evaluation requires some more considerations. According to equation \ref{eq_Qini}, we need for each $t\in [0,1]$ an unbiased estimate of $ATE_t$, where $ATE_t:=E[\tau_x|\hat{\tau}(x)>F_{\hat{\tau}}^{-1}(1-t)]$ denotes the average treatment effect on the share $t$ of highest ranked individuals. To this end, we can apply the principle of stratified estimation in equation \eqref{eq_ate_s} and apply the estimator
\begin{align}
    \hat{ATE}_{t}=p_{H,t}\cdot\hat{\tau}_{H,t}+(1-p_{H,t})\cdot\hat{\tau}_{L,t}, \label{eq_qini_hs}
\end{align} where $p_{H,t}:=P[x\in S_H|\hat{\tau}(x)>F_{\hat{\tau}}^{-1}(1-t)]$ denotes the proportion of individuals in stratum $S_H$, within the share $t$ of individuals with the highest CATE predictions. $\hat{\tau}_{H,t},\hat{\tau}_{L,t}$ denote average treatment effect estimates for the highest ranked individuals within the strata $S_H$, respectively $S_L$. The ATE estimates can either be obtained by simple difference-in-means or covariate adjusted versions of it.

So, the calculation procedure to obtain $\hat{ATE}_{t}$ is straight-forward, however, what is more challenging, is to to obtain the threshold $F_{\hat{\tau}}^{-1}(1-t)$ and the corresponding proportion $p_{H,t}$, required to apply equation \eqref{eq_qini_hs}. This challenge is due to the disproportional sampling, which means that the quantile $F_{\hat{\tau}}^{-1}(1-t)$ of predictions on our HS-sampled data set does not correspond to the quantile $F_{\hat{\tau}}^{-1}(1-t)$ on the whole population. The solution here is to obtain the cut-off $F_{\hat{\tau}}^{-1}(1-t)$, by applying the uplift model $\hat{\tau}(x)$ on a data set with proportions of $S_H$ and $S_L$ matching the proportions in the whole population. This can be done by randomly deleting observations from stratum $S_H$ of the HS-sampled data, such that the proportions of $S_H$ and $S_L$ match the proportions in the whole population or by applying $\hat{\tau}(x)$ on the whole population (because we only need the prediction quantile, we only need the features and not the outcome, so even individuals outside our experimental sample could be used.) Once, $F_{\hat{\tau}}^{-1}(1-t)$ is obtained, it is straight-forward to obtain $p_{H,t}$: One can just calculate on the data, where $F_{\hat{\tau}}^{-1}(1-t)$ was obtained, the proportion of individuals with $\hat{\tau}(x)>F_{\hat{\tau}}^{-1}(1-t)$, which are in stratum $S_H$.

With the above procedure, one can obtain unbiased measures of uplift model performance on HS-sampled data. The principle of variance reduction for ATE estimates, described in Appendix \ref{app_variance_red_techniques} extends directly towards uplift model evaluation. Hence, one can expect to obtain uplift model evaluations on HS-sampled data with far less variance than on randomly sampled data.

\section{Computational experiment}\label{sec_comp}
In this section, we describe how we evaluated the effect of HS-sampling on the ATE and CATE estimation as well as on the uplift model evaluation. Due to the high variance involved in these tasks (see figure \ref{fig:treat_var} for illustration), well-founded statistical considerations had to be made when planning the experiments. To still enable a good reading flow, we decided to describe here the main aspects of our experiment procedure and elaborate on the statistical subtleties in Appendix \ref{app_cate_evaluation}. The following section will describe the experiment procedure we applied on three simulation scenarios and on one real-world data set.  

\subsection{Experiment procedure}
\begin{figure}
\centering
\begin{minipage}{.5\textwidth}
  \centering
  \includegraphics[width=0.95\linewidth]{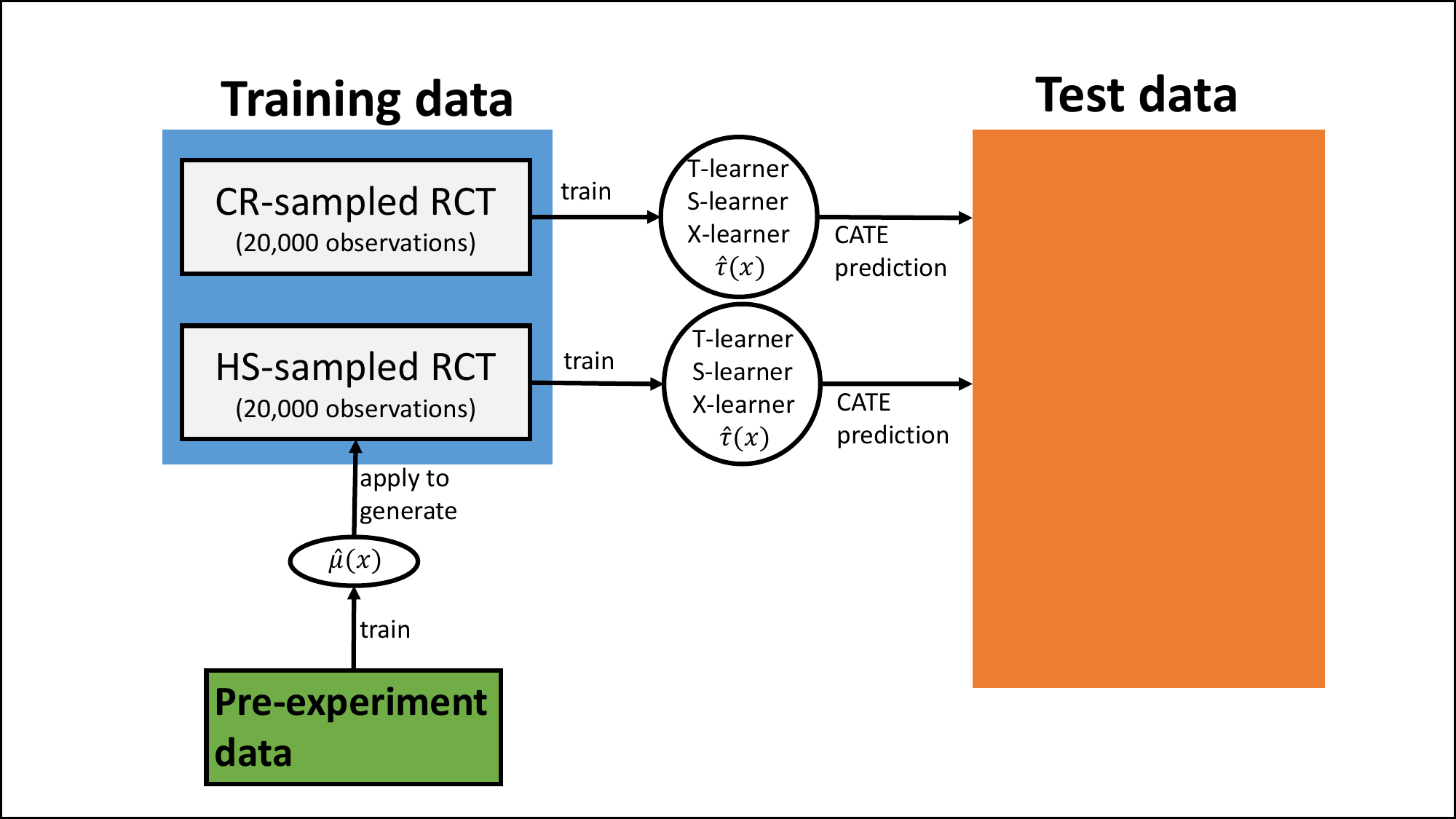}
  \captionof{figure}{Experiment procedure for the uplift model\\ training}
  \label{fig:exp_training}
\end{minipage}%
\begin{minipage}{.5\textwidth}
  \centering
  \includegraphics[width=0.95\linewidth]{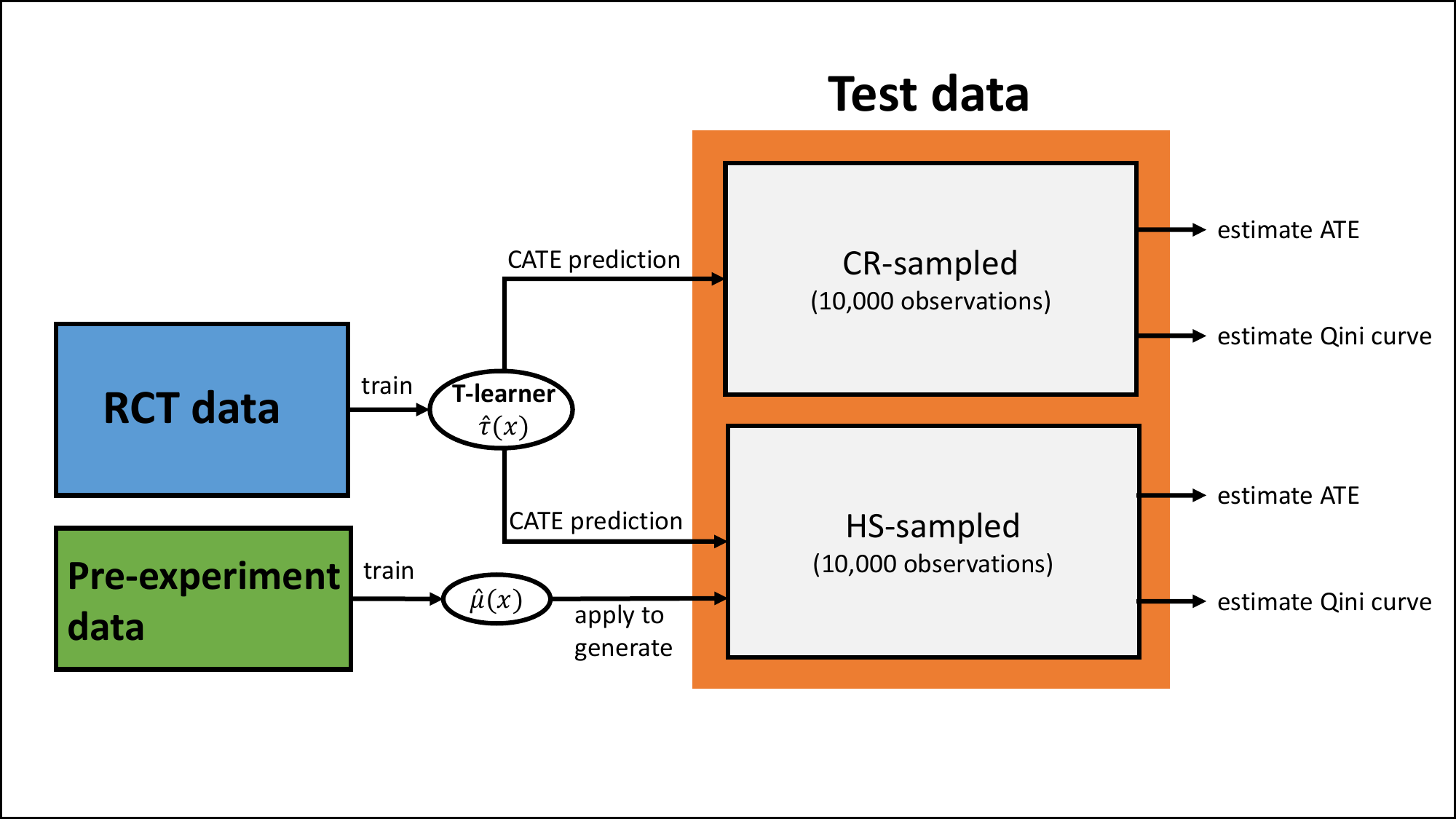}
  \captionof{figure}{Experiment procedure for the ATE and Qini curve estimation}
  \label{fig:exp_test}
\end{minipage}
\end{figure}

Our experiment is divided into one part to assess the effect of HS-sampling on uplift model training and one part to assess the effect of HS-sampling on ATE estimation and uplift model evaluation. These two parts of the experiment are illustrated in figures \ref{fig:exp_training} and \ref{fig:exp_test}. For both parts, an outcome model $\hat{\mu}(x)$ needed to be estimated. We applied random forest throughout and trained it on a set of observations from untreated individuals. As the corresponding data set represents pre-experiment data, the data set was not used in the remainder of the experiment. For the simulation settings, we choose a pre-experimental data set of size 100,000 and for the real-world data set, we choose a pre-experimental part of size 139,000 (1\% of the whole data set). 

In the first part we evaluated the effect of HS-sampling on uplift model training. To this end, we trained six different types of uplift models (T-learner, S-learner, X-learner, each in one version based on random forest and one version based on generalized linear models) on a complete randomly chosen RCT sample and on a HS-sampled RCT set. Thereby, the RCT samples were of size 20,000 throughout. We compared for each type of uplift model the performance when trained on the HS-sampled data with the performance when trained on completely randomly sampled data. As evaluation metric we chose area under the uplift curve (AUQ). Because uplift model evaluation involves much variance \citep{bokelmann2023improving}, we applied the following approaches to obtain reliable results: On real-world data, we used a test set of size 1,000,000. On simulated data, we choose test sets of size 250,000 but further reduced the variance by calculating AUQ based on the (on real-world data unobservabe) CATE values $\tau_x$ (see Appendix \ref{app_cate_evaluation} for details). The process is illustrated in figure \ref{fig:exp_training}. We repeated the experiment 1,000 times to get statistically reliable results and calculate confidence intervals.

In the second part, we examined the effect of HS-sampling on ATE estimation and uplift model evaluation. We start by describing the uplift evaluation procedure. To assess how reliable our uplift model evaluations are, it was sufficient to train a single uplift model (a random forest-based T-learner in our case), and measure how reliable we can evaluate its performance. In contrast to the first part of the computational experiment, we chose relatively small test sets (of size 20,000 throughout) and measured the variance of the Qini curve at each decile. We did this with completely randomly sampled test data and HS-sampled test data. We calculated the variance by repeating the test data sampling and Qini curve calculation part 1,000 times. On each data set used for Qini curve calculation, we also estimated the ATE. The comparison of HS-sampling data-based Qini curve and ATE estimates with complete randomly sampled data-based Qini curve and ATE estimates showed us by how much HS-sampling can reduce the variance (or equivalently increase the precision of such estimates). Since, covariate adjustment is also a common approach for ATE estimate variance reduction (see Appendix \ref{app_covad} for details) and can also be applied to reduce the variance of the Qini curve \citep{bokelmann2023improving}, we decided to include covariate adjustment-based estimates as well in the study. We also examined the combination of covariate adjustment and HS-sampling.       

\subsection{Data}
\subsubsection{Simulated data}\label{sec_sim_scenarios}
We simulated RCT data, with a treatment proportion of $p=0.5$. We thereby simulated three different scenarios:

Scenario 1
\begin{align*}
    x_{ji}&\sim N(0,1)\\
    y_i&\sim Bern(\sigma(x_{1i}+0.5\cdot x_{2i}+x_{3i}\cdot x_{4i}-4+0.1\cdot w_i))
\end{align*}

Scenario 2
\begin{align*}
    x_{ji}&\sim N(0,1)\\
    y_i&\sim Bern(\sigma(x_{1i}^2+0.5\cdot x_{2i}+x_{3i}\cdot x_{4i}-7+(1.1+x_{5i})\cdot w_i))
\end{align*}

Scenario 3
\begin{align*}
    x_{ji}&\sim N(0,1)\\
    y_i&\sim Bern(\sigma(0.1\cdot e^{x_{1i}}+0.5\cdot x_{2i}^3+x_{3i}-7+(0.1+x_{5i}\cdot x_{6i})\cdot w_i))
\end{align*}

\subsubsection{Real-world data}
As real-world data, we used the Criteo large scale benchmarking data set.\cite{diemert2021large}. It contains RCT data with 13,979,592 observations and a treatment proportion of $p=0.85$. The data comes from an online marketing application. We chose conversion as our target $y$ of interest. The reason why we choose the Criteo data set is its size, which allowed us to perform statistically meaningful experiments.

\subsection{Results}

\subsubsection{Uplift model training}
The results of the experiment to evaluate the effect of HS-sampling on uplift model training are provided in table \ref{tab_cate_train}. Uplift models trained on HS-sampled data performed almost always better then uplift models trained on completely randomly sampled RCT data. Especially on the real-world Criteo data set do we notice a huge performance improvement by HS-sampling with the HS-sampling data-based S-learner being 57.41\% (in the GLM version) and 24.32\% (in the random forest version) better than the respective versions trained on completely randomly sampled data.

Only in 3 out of 24 comparisons was the uplift model trained on completely randomly sampled data then the respective uplift model trained on HS-sampled data. In all 3 cases, these were GLM-based uplift models. We found that the uplift model versions, where completely randomly sampled data served better than HS-sampled data were in no scenario among the best performing versions. The best models per scenario were always trained on HS-sampled data. Specifically these were: the random forest-based X-learner with an AUQ of 0.0764 on simulation scenario 1, the random forest-based X-learner with an AUQ of 0.021 on simulation scenario 2, the random forest-based X-learner with an AUQ of 0.0026 on simulation scenario 3 and the GLM-based S-learner with an AUQ of 0.0099 on the criteo data set.

In summary, we can see clear evidence of a beneficial effect of HS-sampling on uplift model performance. We would attribute the results where uniformly sampled data led to better results than HS-sampled data rather to the unsuitability of the respective uplift model versions in these settings than on deficiencies in the HS-sampling approach. 

\begin{threeparttable}
\caption{Improvement of uplift model performance}\label{tab_cate_train}
\begin{tabular*}{\textwidth}{p{0.13\textwidth}| p{0.12\textwidth}p{0.12\textwidth}p{0.12\textwidth}p{0.12\textwidth}p{0.12\textwidth}p{0.12\textwidth}}
    \toprule
    \textbf{data}  & T-learner (LR) & S-Learner (LR) & X-Learner (LR)  & T-learner (RF) & S-Learner (RF) & X-Learner (RF) \\
    \hline\hline
    scenario 1  & -0.22\% $[-0.35;-0.09]$ & 0.2\% [0.16;0.23] & -0.09\% $[-0.16;-0.03]$ & 0.1 $[0.04;0.16]$ & 0.08\% $[0.02;0.13]$ & 0.13\% $[0.09;0.18]$ \\
    \hline 
    scenario 2  & 1.56\% [1.39;1.73] & 1.67\% [1.56;1.78] & 0.78\% [0.66;0.9] & 1.96\% [1.87;2.05] & 1.73\% [1.65;1.81] & 1.69\% [1.62;1.76] \\
    \hline
    scenario 3  & 15.52\% [14.81;16.22] & 0.89\% $[-0.21;1.98]$ & 4.65\%  [3.52;5.77] & 9.07\% [8.64;9.51] & 5.9\% [5.52;6.27] & 11.5\% [11.0;12.0] \\
    \hline
    Criteo  & -1.37\% $[-2.26;-0.48]$ & 57.41\% [46.9;67.91]  &  25.66\% [23.18;28.14] & 7.17\% [5.96;8.38] & 24.32\% [21.56;27.09] & 16.42\% [14.34;18.5] \\
    \hline
     
\end{tabular*}

\begin{tablenotes}[para,flushleft]
   \small
   The table shows the percentage increase in area under the Qini curve of a model trained on HAS-sampled RCT data, compared to a model trained on completely random sampled RCT data. We provide the mean increase as well as a 95\% confidence interval. By following the HS-sampling procedure in section \ref{sec_hs_sampling} with parameter adjustment (see section \ref{sec_hs_robust}), we obtained the following sampling parameters:\\
   scenario 1: $p_H=0.3, R_H=1.53$\\
   scenario 2: $p_H=0.1, R_H=2.9$\\
   scenario 3: $p_H=0.05, R_H=4.4$\\
   Criteo: $p_H=0.1, R_H=3.4$
   
   \end{tablenotes}

\end{threeparttable}

\subsubsection{ATE estimation and CATE model evaluation}
The results for the ATE estimation and the CATE model evaluation are provided in Table \ref{tab_ate_cate_eval}. In all considered simulation scenarios and on the real-world data, we see a notable variance reduction of the ATE estimator by using HS-sampled data instead of randomly sampled data. The most impressive variance reductions were achieved in simulation scenario 3 where HS-sampling lead to a variance reduction of 59.0\% and on the Criteo data, where it lead to a variance reduction of 54.2\%. We can also see, that there is additional variance reduction if HS-sampling is combined with post-experiment covariate-adjustment. For simulation scenario 3, the combination resulted in a variance reduction of 67.5\% and on the Criteo data set of 57.2\%. 

The results for the variance reduction of the Qini curve were similar in magnitude. Overall, the results clearly showed that HS-sampling has the potential to significantly reduce the variance of ATE estimation and uplift model evaluation.

\begin{threeparttable}
\caption{Variance reduction for ATE and Qini curve estimates}\label{tab_ate_cate_eval}
\begin{tabular*}{\textwidth}{p{0.13\textwidth}| p{0.12\textwidth}p{0.12\textwidth}p{0.12\textwidth}p{0.12\textwidth}p{0.12\textwidth}p{0.12\textwidth}}
    \toprule
    \textbf{data}  & $\hat{ATE}_{CA}$ & $\hat{ATE}_{HS}$ & $\hat{ATE}_{HSCA}$  & $Qini_{CA}$ & $Qini_{HS}$ & $Qini_{HSCA}$ \\
    \hline\hline
    scenario 1  & 23.6\% & 15.9\% & 30.2\% & 19.5-23.6\% & 15.9-34.8\% & 30.2-46.8\%  \\
    \hline 
    scenario 2  & 40.1\% & 40.2\% & 49.9\% & 31.8-40.4\% & 19.3-41.3\% & 32.0-50.4\% \\
    \hline
    scenario 3  & 48.5\% & 59.0\% & 67.5\% & 48.5-57.1\% & 59.0-74.0\% & 67.4-82.2\% \\
    \hline
    Criteo  & 10.9\% & 54.2\% & 57.2\% & 9.1-10.9\% & 52.6-66.0\% & 56.5-83.4\% \\
    \hline
     
\end{tabular*}

\begin{tablenotes}[para,flushleft]
   \small
   By following the HS-sampling procedure in section \ref{sec_hs_sampling} with parameter adjustment (see section \ref{sec_hs_robust}), we obtained the following sampling parameters:\\
   scenario 1: $p_H=0.3, R_H=1.53$\\
   scenario 2: $p_H=0.1, R_H=2.9$\\
   scenario 3: $p_H=0.05, R_H=4.4$\\
   Criteo: $p_H=0.1, R_H=3.4$
   \end{tablenotes}

\end{threeparttable}

\section{Discussion}\label{sec_disc}
In this paper, we proposed heteroskedasticity-aware stratified sampling, to choose a suitable sample for an RCT. This rather simple sampling procedure requires binary outcomes and the existence of pre-experimental data from customers not having received treatment. The basic idea is to divide the customer base into two strata $S_H,S_L$ according to their expected high/low outcome and then to sample customers from $S_H$ with an excessive proportion in the RCT sample. Because statistical theory indicates that observations from $S_H$ contain more noise than observations from $S_L$, this excessive proportion in the RCT sample is necessary to reliably estimate treatment effects of them. According to statistical theory, the HS-sampling scheme can be expected to achieve a variance reduction in the ATE estimation, compared to ATE estimation on completely randomly sampled RCT data. The same statistical considerations also suggest that HS-sampling leads to a variance reduction in the estimation of the Qini curve for uplift models. As HS-sampling generally improves treatment effect estimation, we also expected to see improvement in uplift model performance when trained on HS-sampled data.

Our computational experiment confirmed the theoretical considerations. On all three simulation scenarios as well as on the real-world data, we saw a significant variance reduction of the ATE and Qini curve estimators by HS-sampling. The variance reduction by HS-sampling was comparable to the variance reduction by the established covariate adjustment variance reduction method. It is important to highlight, that these two variance reduction methods affect different sources of variance and hence a combination of the methods reliably leads to more variance reduction then when applying one of the methods alone. Our computational experiment also clearly showed the performance gain of uplift models, when trained on HS-sampled data. Hence, our paper provides sound evidence for the usefulness of HS-sampling, when collecting RCT data. 

The definition of the HS-sampling parameters (threshold $p_H$ and oversampling ratio $S_H$) depend on the predictions of an outcome model, trained on pre-experimental data. As such predictions necessarily contain some degree of error, we conducted robustness considerations of our HS-sampling procedure. Instead of choosing the parameters which would be deemed optimal based on the outcome model predictions, we suggest to use adjusted parameter values. In our computational experiment, we adjusted $p_H$ and $S_H$, by upward respectively downward shifting both by one quarter. This was a rather pragmatic approach and the value of one quarter was rather arbitrarily chosen. However, we expect that it might be difficult to define an adjustment approach on more sophisticated theoretical considerations. This is because the demand for adjustment is due to prediction errors of the outcome models (in particular its degree of over-fitting) and this problem might be different from case to case. In any case, it is possible to choose a lower oversampling rate than would be deemed optimal by the outcome model and thereby prevent any potential adverse effects of the HS-sampling approach.

There are ways in which our proposed HS-sampling procedure could potentially be extended: It would be possible to define more than two strata and calculate their sampling proportions, based on the respective expected outcome variance. This might lead to even more variance reduction than our approach with two strata. But we expect a decreasing marginal gain when increasing the number of strata. HS-sampling achieves its beneficial effect due to the variance heterogeneity between the strata. Defining two strata in the way which we proposed already leads to strong variance heterogeneity between these strata. It is unlikely that more splits will lead to additional variance heterogeneity in the same magnitude. Moreover, statistical properties and applicational aspects become more challenging with a growing number of strata: With more strata, the number of individuals per stratum decreases and so the outcome variance estimates based on which the HS-sampling parameters are chosen become more unreliable and the procedure to calculate the Qini curve also becomes more complicated as each new stratum $S$ requires to estimates ($p_{S,t},\hat{\tau}_{S,t}$) for the values $t \in [0,1]$ (see section \ref{sec_cate_eval}). Hence, with regard to these statistical and applicational aspects we expect our suggested solution with two strata to be already a suitable one for practice.      

\appendix
\section{Variance reduction techniques}\label{app_variance_red_techniques}
\subsection{Difference in means with random sampling}\label{app_dim}
The difference in means estimator can be written as
\begin{align}
    \hat{ATE}&=\frac{1}{N_w}\sum_{w_i=1} y_{i}-\frac{1}{N_{\bar{w}}}\sum_{w_i=0} y_{i}\notag\\
    &=\frac{1}{N}\sum W^{p}_iy_i\notag\\
    &=\frac{1}{N}\sum W^{p}_i(\mu_{x_i}+w_i\cdot\tau_{x_i})+\frac{1}{N}\sum W^{p}_i\varepsilon_i\notag\\
    &=\frac{1}{N}\sum\zeta_i+\frac{1}{N}\sum \bar{\varepsilon}_i.\label{eq_ate_var_comp}
\end{align} Thereby, we use the operator $W^{p}_i:=\begin{cases}
			\frac{1}{p}, & \text{if $w_i=1$}\\
            -\frac{1}{1-p}, & \text{if $w_i=0$}
\end{cases}$ and define $\zeta_i:=W^{p}_i(\mu_{x_i}+w_i\cdot\tau_{x_i})$ and $\bar{\varepsilon}_i:=W^{p}_i\varepsilon_i$.

The variance of the ATE estimator with random sampling can then be decomposed into
\begin{align*}
    Var[\hat{ATE}]=\frac{Var[\zeta]}{N}+\frac{Var[\bar{\varepsilon}]}{N}.
\end{align*}

\subsection{Stratified estimation with proportional sampling}\label{app_strat}
The stratified estimator, with two strata $S_H,S_L$ and proportion $p_H$ of individuals from strata $S_H$ in the whole population, is given by
\begin{align*}
    \hat{ATE}_{S}=p_{H}\cdot\hat{\tau}_{H}+(1-p_H)\cdot\hat{\tau}_{L}
\end{align*}

It is easy to show, that this estimator is unbiased.
\begin{align*}
    E[\hat{ATE}_{S}]&=p_{H}\cdot E[\hat{\tau}_{H}]+(1-p_H)\cdot E[\hat{\tau}_{L}]\\
    &=P[x\in S_H]\cdot E[\tau_x|x\in S_H]+(1-P[x\in S_H])\cdot E[\tau_x|x\in S_L]\\
    &=E[\tau_x]\\
    &=ATE
\end{align*}

Next, we analyse its variance. Therefore, we first examine the variance of the difference-in-means estimators $\hat{\tau}_H,\hat{\tau}_L$ on the strata $S_H,S_L$. We can apply a variance decomposition on both strata, in the same way as in the last sub-section. By noting that the sample size on $S_H$ is $p_H\cdot N$ and the sample size on $S_L$ is $(1-p_H)\cdot N$, we can derive   
\begin{align*}
    Var[\hat{\tau}_{H}]&=\frac{Var[\zeta|S_H]+Var[\bar{\varepsilon}|S_H]}{p_H\cdot N}\\
    Var[\hat{\tau}_{L}]&=\frac{Var[\zeta|S_L]+Var[\bar{\varepsilon}|S_L]}{(1-p_H)\cdot N}.
\end{align*} Hence, it follows
\begin{align}
    Var[\hat{ATE}_{S}]&=p_H^2\cdot Var[\hat{\tau}_{H}]+(1-p_H)^2\cdot Var[\hat{\tau}_{L}]\label{eq_var_strat_alt}\\
    &=\frac{\left(p_H\cdot Var[\zeta|S_H]+(1-p_H)\cdot Var[\zeta|S_L]\right)+\left(p_H\cdot Var[\bar{\varepsilon}|S_H]+(1-p_H)\cdot Var[\bar{\varepsilon}|S_L]\right)}{N}\notag\\
    &=\frac{E\left[Var[\zeta|S]\right]+E\left[Var[\bar{\varepsilon}|S]\right]}{N}\notag\\
    &=\frac{E\left[Var[\zeta|S]\right]}{N}+\frac{Var[\bar{\varepsilon}]}{N}\notag
\end{align}

Hence, stratification removes the variance component $\frac{Var\left[E[\zeta|S]\right]}{N}$ from the variance of the random sampling-based difference in means ATE estimator. Notably, the variance component $\frac{Var[\bar{\varepsilon}]}{N}$ is not affected. 

\subsection{Stratified estimation with optimal allocation sampling}\label{app_strat_opt}
If we are flexible in the choice of sample sizes $N_H,N_L$ per stratum (under the condition $N_H+N_L=N$), the variance of the stratified estimator becomes
\begin{align}
    Var[\hat{ATE}_S]=&p_H^2\cdot\frac{1}{N_H}\cdot \left(\frac{Var[y|w=1,S_H]}{p}+\frac{Var[y|w=0,S_H]}{1-p}\right)\notag\\
    &+(1-p_H)^2\cdot\frac{1}{N-N_H}\cdot \left(\frac{Var[y|w=1,S_L]}{p}+\frac{Var[y|w=0,S_L]}{1-p}\right)\notag\\
    =&p_H^2\cdot\frac{1}{N_H}\cdot V_H+(1-p_H)^2\cdot\frac{1}{N-N_H}\cdot V_L,\label{eq_var_strat_unprop}
\end{align} with
\begin{align*}
    V_i:=\frac{Var[y|w=1,S_i]}{p}+\frac{Var[y|w=0,S_i]}{1-p},
\end{align*} for $i=H,L$.

To find the optimum sampling proportion $\frac{N_H}{N}$, one can simply take the first two derivatives
\begin{align*}
    \frac{\partial Var[\hat{ATE}_S]}{\partial N_H}&=-\frac{V_H\cdot p_H^2}{N_H^2}+\frac{V_L\cdot (1-p_H)^2}{(N-N_H)^2}\\
    \frac{\partial^2 Var[\hat{ATE}_S]}{(\partial N_H)^2}&=2\left(\frac{N_H\cdot p_H^2}{N_H^3}+\frac{V_L\cdot (1-p_H)^2}{(N-N_H)^3} \right).
\end{align*} The first derivative gets zero for
\begin{align*}
    N_H=N\cdot\frac{p_H\cdot\sqrt{N_H}}{p_H\cdot\sqrt{N_H}+(1-p_H)\cdot\sqrt{V_L}}.
\end{align*} As the second derivative is positive for this choice of $N_H$, it leads to the minimum variance of the estimator.

The corresponding estimator variance for optimal allocation becomes
\begin{align*}
    Var[\hat{ATE}_{Sopt}]=\frac{(p_H\cdot\sqrt{V_H}+(1-p_H)\cdot\sqrt{V_L})^2}{N}
\end{align*} 

The variance reduction compared to the stratified ATE estimation approach with proportional sampling is by allocating more observations to the stratum $S_i$, where the variance $V_i$ is higher. So, the optimal allocation has a variance reduction effect on the components responsible for the difference between $V_H=Var[\zeta|S_H]+Var[\bar{\varepsilon}|S_H]$ and $V_L=Var[\zeta|S_L]+Var[\bar{\varepsilon}|S_L]$. We would expect $Var[\bar{\varepsilon}|S_i]$ to play the biggest role in the difference between $V_H$ and $V_L$, as (1) we would expect that in most practical applications most of the variance in the outcome can not be explained by the features and is therefore due to $\varepsilon$ and (2) the stratification takes place by estimates of $\mu_x$, so the variance of $\zeta$ within each stratum should be limited.       

\subsection{Covariate adjustment}\label{app_covad}
Covariate adjustment is a variance reduction procedure, which can be applied after the experiment is carried out. The idea is to adjust for differences in the outcomes between the intervention and the control group, which are not due to the treatment, but rather due to a random unbalance in the feature distribution between the intervention and the control group.  

A traditional method for covariate adjustment in online experiments is CUPED, where a linear regression model is used to adjust for feature unbalance between intervention and control group.\citep{deng2013improving} More recently, machine learning methods for covariate adjustment have been proposed.\citep{guo2021machine,hosseini2019unbiased,cohen2020no,jin2023toward} Thereby \citet{jin2023toward} suggests a covariate adjustment procedure, which asymptotically leads to the optimal variance reduction, as long as the applied machine learning algorithms are consistent. They suggest to estimate the ATE by  
\begin{align*}
    \hat{ATE}_{CV}&=\frac{1}{K}\hat{\tau}_{k}\text{ with}\\
    \hat{\tau}_{k}&=\frac{1}{n_k}\sum_{i\in D^{(k)}}(\hat{\mu}_1(x_i)-\hat{\mu}_0(x_i))+\frac{1}{n_{k,t}}\sum_{w_i=1,i\in D^{(k)}}(y_i-\hat{\mu}_1(x_i))+\frac{1}{n_{k,c}}\sum_{w_i=0,i\in D^{(k)}}(y_i-\hat{\mu}_0(x_i)).
\end{align*} Thereby, the data is split into $K$ folds and the total estimate $\hat{\tau}$ is calculated as an average of the $K$ estimates $\hat{\tau}_k$ on the individual folds. For each fold $k$, $n_k,n_{k,c},n_{k,t}$ denote the number of total observations, respectively observations in the control group, respectively observations in the treatment group. The machine learning models 
$\hat{\mu}_0,\hat{\mu}_1$, which are used for the covariate adjustment, are trained to predict the outcomes of the untreated, respectively treated. For each fold $k$, they are trained on the complete data except for fold $k$ and then applied to predict on fold $k$. This fold-wise estimation procedure leads to unbiased estimates and is known under the name \textit{cross-fitting}.\citep{chernozhukov2018double}

\citet{jin2023toward} derived that their covariate adjustment procedure asymptotically leads to the highest variance reduction, as long as the machine learning model predictions $\hat{\mu}_0(x),\hat{\mu}_1(x)$ converge to their prediction targets $\mu_{x}$, respectively $\mu_x+\tau_x$ with growing sample size. So, the optimal variance reduction would be achieved, if the machine learning model predictions would be replaced by their respective targets in the above equations. If adjustment is done with the actual targets instead of the machine learning model predictions, the cross-fitting procedure is unnecessary, because the model training part, which could lead to a bias in the estimation procedure, is removed. Hence, an estimator of the form
\begin{align}
    \hat{\tau}&=\frac{1}{N}\sum(\mu_{x_i}+\tau_{x_i}-\mu_{x_i})+\frac{1}{N_w}\sum_{w_i=1}(y_i-\mu_{x_i}-\tau_{x_i})-\frac{1}{N_{\bar{w}}}\sum_{w_i=0}(y_i-\mu_{x_i})\notag\\
    &=\frac{1}{N_w}\sum_{w_i=1}(y_i-\mu_{x_i}-\frac{N_{\bar{w}}}{N}\cdot\tau_{x_i})-\frac{1}{N_{\bar{w}}}\sum_{w_i=0}(y_i-\mu_{x_i}-\frac{N_{\bar{w}}}{N}\cdot\tau_{x_i})\notag\\
    &=\frac{1}{N_w}\sum_{w_i=1}(y_i-\Phi(x_i))-\frac{1}{N_{\bar{w}}}\sum_{w_i=0}(y_i-\Phi(x_i))\label{eq_dr},
\end{align} where $\Phi(x_i):=\mu_{x_i}+(1-p)\cdot\tau_{x_i}$ with $p$ being the proportion of treated individuals, yields the optimal variance reduction possible with covariate adjustment. Of course, in real-world data, it would not be possible to apply this estimator, because $\mu_x$ and $\tau_x$ are not observable. However, in this study we perform simulation settings, in which estimation equation \eqref{eq_dr} can directly be applied. This is useful, because it provides us the maximum variance reduction, which could be achieved by covariate adjustment. In addition, equation \eqref{eq_dr} is usefull, because it shows the difference in the variance reduction principle between covariate adjustment and our proposed heteroskedasticity aware stratified sampling. 

Covariate adjustment not only reduces the variance of ATE estimates, but also reduces the variance of uplift evaluation metrics like the Qini curve. Following our paper \citep{bokelmann2023improving}, adjusted outcomes of the form $y_i-\Phi(x_i)$ could also be used to reduce the variance of the Qini curve.

To see, which variance components of the ATE estimator covariate adjustment reduces, we can replace the original outcome by the adjusted outcome $y_i-\Phi(x_i)$ in equation \eqref{eq_ate_var_comp}. This yields
\begin{align*}
    \hat{ATE}_{CV}&=\frac{1}{N}\sum(\zeta_i-\Phi(x_i))+\frac{1}{N}\sum \bar{\varepsilon}_i\\
    &=\frac{Var[\zeta-\Phi(x)]}{N}+\frac{Var[\bar{\varepsilon}]}{N}.
\end{align*} Notably, there is a reduction of the component $\frac{Var[\zeta]}{N}$, but no reduction of $\frac{Var[\bar{\varepsilon}]}{N}$ compared to the difference-in-means estimator. 

\subsection{Pre-experiment covariate balancing}
Completely randomized treatment allocation can lead to an inbalance in the feature distribution between the treatment and the control group. This possibility of inbalance is a factor of variance in the ATE estimation and can also negatively effect CATE estimation. 

There is a wide range of methods to restrict the randomness in the treatment allocation in order to guarantee a more balanced feature distribution in the treatment and control group. Here, we only provide an overview of according research in recent years. Detailed description of the corresponding methods can be found in the respective references. Traditional statistical methods to achieve covariate balance include blocking \citep{greevy2004optimal,higgins2016improving}, matching \citep{imai2008variance} and rerandomization \citep{li2018asymptotic}. More recently, kernel-based treatment allocation \citep{kallus2018optimal}, minimum spanning trees \citet{arbour2021efficient} and Gram-Schmidt walk-based treatment allocation \citep{harshaw2023balancing} were introduced.

The likelihood of covariate imbalance affects the distribution of $\zeta^{(1)}=\frac{1}{N_w}\sum_{w_i=1}(\mu_{x_i}+\tau_{x_i}-\mu-\tau)$ and $\zeta^{(0)}=\frac{1}{N_{\bar{w}}}\sum_{w_i=0}(\mu_{x_i}-\mu)$. By ensure that the feature distribution in the treatment and the control group is as similar as possible to the feature distribution in the whole population, the covariate balancing techniques reduce the variance of $\frac{Var[\bar{\zeta}]}{N}$. The effect on the variance of $\frac{Var[\bar{\varepsilon}]}{N}$ is marginal. 

\section{HS-sampling robustness considerations}\label{app_robustness}
\subsection{Choice of the oversampling ratio $R_H$}
We need to examine under which conditions and in which way the chosen $R_H(\hat{Q}_V)$ deviates from the optimal $R_H(Q_V)$. Using equation \eqref{eq_opt_proportion}, we can see that $R_H(\hat{Q}_V)$ monotonically increases in $\hat{Q}_V$, so we need to answer when $\hat{Q}_V$ heavily overestimates or underestimates the true variance quotient $Q_V$. The variance quotient can be written as
\begin{align*}
    Q_V=\frac{V_H}{V_L}=\frac{E[\mu_x|S_H]\cdot (1-E[\mu_x|S_H])}{E[\mu_x|S_L]\cdot (1-E[\mu_x|S_L])}.
\end{align*} To obtain the estimate $\hat{Q}_V$, we use predictions $\hat{\mu}(x)$ as a proxy for $\mu_x$. To see, when this leads to upward or downward biased estimates, it is useful to write the predictions as 
\begin{align}
    \hat{\mu}(x)=\mu_{x}+\tilde{\varepsilon},\label{eq_dist_omodel}
\end{align} with a certain error term $\tilde{\varepsilon}$, for which we can assume $E[\tilde{\varepsilon}]=0$. Hence, the outcome variance of a stratum $S_i$ gets overestimated if $E[\tilde{\varepsilon}|S_i]>0$ and gets underestimated if $E[\tilde{\varepsilon}|S_i]<0$. So, we need to analyze when these cases occur. 

Due to the definitions of the strata we can write
\begin{align*}
    E[\tilde{\varepsilon}|S_H]&=E[\tilde{\varepsilon}|\mu_x+\tilde{\varepsilon}>F^{-1}_{\hat{\mu}}(1-p_H)]\text{ and}\\
    E[\tilde{\varepsilon}|S_L]&=E[\tilde{\varepsilon}|\mu_x+\tilde{\varepsilon}\leq F^{-1}_{\hat{\mu}}(1-p_H)].
\end{align*} Clearly, there is a form of selection, which will tend to put predictions with $\tilde{\varepsilon}>0$ in stratum $S_H$ and predictions with $\tilde{\varepsilon}<0$ in stratum $S_L$. So, the stratum definition is one relevant factor, which always tends to upward bias the estimate $\hat{Q}_V$. But it is not the only relevant factor.

The second factor is related to the model building of $\hat{\mu}(x)$. Namely, it is the degree to which $\hat{\mu}(x)$ over-fits. When building a predictive model, complexity restrictions lead to a shrinkage of predictions $\hat{\mu}(x_i)$ towards the unconditional expected value $E[\mu_x]$. This would cause a tendency of $\tilde{\varepsilon}$ to become negative for $\hat{\mu}(x)>E[\mu_x]$ and positive for $\hat{\mu}(x)<E[\mu_x]$. Hence, this model complexity restrictions will tend to downward bias the estimate $\hat{Q}_V$ and thus counteract the upward bias due to the stratum selection. However, this tendency of shrinkage to the mean depends on the chosen complexity of $\hat{\mu}(x)$. If a high model complexity is chosen (i.e. $\hat{\mu}(x)$ over-fits the training data) the shrinkage effect diminishes. Hence, the higher the chosen model complexity, the higher the potential upward bias of $\hat{Q}_V$ will be.

As a means to prevent choosing a wrong oversampling ratio $R_H$, we decided to apply two kinds of adjustment measures. First, we made one statistical consideration about the choice of $p_H$: If $p_H$ gets close to 1, the estimate of $\hat{V}_H$ necessarily becomes less reliable, because there are only few observations on which to base this estimate. Accordingly, we decided to shift the value of $p_H$ more towards 0.5. Hence, we decided to use adjusted values of $p_H$, defined by
\begin{align*}
    p_H^{ad}:=p_H+\frac{1}{4}p_H.
\end{align*} This leads to proportions of $S_H$ closer to 0.5, because for our low outcome rate scenarios $E[y]<<0.5$, the estimated optimal proportion is also mostly far below 0.5. The choice to increase $p_H$ by a quarter was rather arbitrary. The second kind of adjustment concerns $R_H$. As we would always expect that $\hat{V}_H$ is upward biased, we decided to shift the optimal oversampling ratio based on the estimated variance quotient $\hat{Q}_V$ towards 1, by taking the weighted average
\begin{align*}
    R^{ad}_H:=\frac{3}{4}\cdot R_{H}(\hat{Q}_V)+\frac{1}{4}\cdot 1.
\end{align*} Our choice of the weighting factor $1/4$ is again rather arbitrary. In practice, one can also shift $R_H$ more towards 1, if one expects $\hat{Q}_V$ to highly overestimate $Q_V$. In the next section, we investigate the potential problem of HS-sampling due to the overestimation of $Q_V$. The results show, that only if $\hat{\mu}(x)$ is strongly overfitted, a relevant problem with the choice of $R_H$ can occur. The results also show, that our adjustment $R^{ad}_H$ provides a robust solution even in this case. 

\subsection{Robustness simulation} 
We simulated predictions of an outcome model by
\begin{align}
    \hat{\mu}(x)&=\alpha\cdot \tilde{\mu}(x)+(1-\alpha)\cdot E[\mu_x]\label{eq_sim_predictions} \text{ with}\\
    \tilde{\mu}(x)&\sim\beta(\nu\cdot\mu_x,\nu\cdot (1-\mu_x)),\notag
\end{align} where $\beta$ denotes the $\beta$-distribution and $\nu$ denotes a parameter of prediction quality. The higher $\nu$, the more accurate the predictions become. $\alpha$ is a shrinkage factor, with $\alpha=1$ being the case of extreme over-fitting, while the ideal shrinkage (leading to minimum mean squared error) toward the mean would be with $\alpha=Cor(\tilde{\mu}(x),\mu_x)\cdot\frac{\sigma(\mu_x)}{\sigma(\tilde{\mu}(x))}$, where $\sigma$ denotes the standard deviation. We took the idea to simulate outcome model predictions for binary outcomes with $\beta$-distribution from \citet{fernandez2022causal}. There simulation scenario corresponds to our $\alpha=1$ scenario.

We used the three simulation scenarios described in section \ref{sec_sim_scenarios} to simulate the true conditional expectations $\mu_x$. We simulated outcome model predictions using equation \eqref{eq_sim_predictions} for an over-fitting scenario ($\alpha=1$) and an optimal tuning scenario ($\alpha=Cor(\tilde{\mu}(x),\mu_x)\cdot\frac{\sigma(\mu_x)}{\sigma(\tilde{\mu}(x))}$). We applied values for $\nu$ ranging from 0.2 to 200. The $\nu$ values affected the prediction accuracy, but do not serve as an interpretable measure for prediction accuracy on their own. Hence, for simulated predictions, we always calculated the accuracy measure
\begin{align*}
    1-\frac{MSE(\hat{\mu}(x),\mu_x)}{Var[\mu_x]},
\end{align*} involving the mean squared error divided by the variance of the true conditional expected values. If a model is not worse than the trivial prediction by the outcome average, this quality measure takes values between 0 and 1, where 1 would be perfect prediction accuracy and 0 would be the prediction accuracy of a trivial model. This accuracy measure was chosen for the x-axis of our plots for the HS-sampling evaluation. 

\begin{figure}
\centering
\centering
  \includegraphics[width=.9\linewidth]{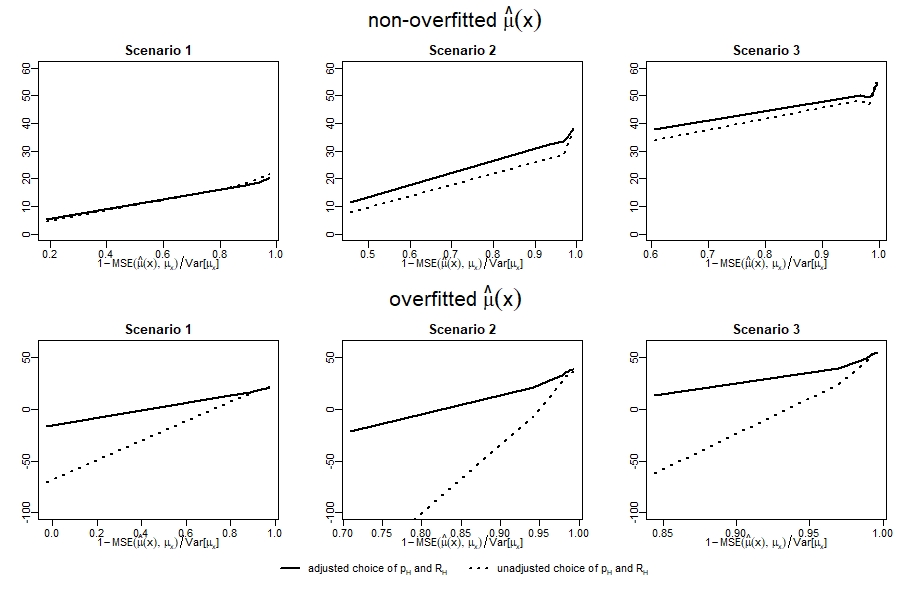}
  \captionof{figure}{}
  \label{fig:robust_sim}
\end{figure}

For a given prediction accuracy of the outcome model, we were interested in the resulting variance reduction by HS-sampling. The results for the three simulation scenarios are provided in figure \ref{fig:robust_sim}. We evaluated HS-sampling based on the estimated optimally chosen values $p_H$ and $R_H(\hat{Q}_V)$ and the respected adjusted values $p^{ad}_H$ and $R^{ad}_H(\hat{Q}_V)$. 

Using the adjusted values the HS-procedure is far more robust against choosing a too high oversampling ratio. Accordingly, the variance reduction almost never becomes negative, even for highly over-fitted outcome models with low accuracy. This is a notable advantage compared to the unadjusted HS-sampling parameters. Even for very accurate outcome models (when the accuracy measure is close to 1), there is almost no loss in variance reduction, when applying the suggested adjusted parameters.  

\section{Statistical considerations in the computational experiment}\label{app_cate_evaluation}

\subsection{Calculation of the Qini curve and the AUQ}
The Qini curve involves estimation of $ATE_{t}\cdot t$ for $t\in [0,1]$. On randomly sampled RCT data, it is straight-forward to estimate $ATE_{t}$: One takes the $(1-t)$-quantile $F_{\hat{\tau}}^{-1}(1-t)$ of the predictions $\hat{\tau}$ on the test set. One then takes all the observations $(x_i,w_i,y_i)$, for which $\hat{\tau}(x_i)\geq F_{\hat{\tau}}^{-1}(1-t)$ to estimate $ATE_{t}$ by a simple difference in means between treated and untreated individuals. 

To quantify the performance in a single value, the area under the Qini curve 
\begin{align*}
    AUQ&=\int_{0}^{1}ATE_{t}\cdot t dt
\end{align*} is a plausible choice.\citep{devriendt2020learning} Using that for each $t\in [0,1]$, it holds $ATE_t=\tau_x$ for $x \text{ with } F_{\hat{\tau}}(\hat{\tau}(x))=t$, we obtain
\begin{align*}
    AUQ=E[F_{\hat{\tau}}(\hat{\tau}(x))\cdot\tau_{x}].
\end{align*} This equation helps us to calculate the AUQ on a test set. On simulation data, we observe $\tau_x$ and hence we are able to estimate the AUQ directly by
\begin{align*}
    \hat{AUQ}=\frac{1}{N}\sum_{i=1}^{N} F_{\hat{\tau}}(\hat{\tau}(x_i))\cdot\tau_{x_i}
\end{align*} On real-world data, it is possible to split the data into $T$ parts, according to the $t/T$ quantiles of $\hat{\tau}$ ($t\in \{0,1,...,T\}$) and then estimate the AUQ via
\begin{align*}
    \hat{AUQ}=\frac{1}{T}\sum_{t=0}^{T}\frac{T-t}{T}\hat{ATE}_{t/T}. 
\end{align*} Compared to the estimation procedure on simulated data, the real-world data procedure suffers from higher variance due to the noise in the estimate $\hat{ATE}_{t/T}$.

\subsection{Covariate adjustment for uplift model evaluation}
\citet{bokelmann2023improving} investigated on the possibility of covariate adjustment for uplift model evaluation. Similar to the case of ATE estimation, it would lead to the strongest variance reduction, if the outcome is adjusted to $y-\Phi(x)$ with $\Phi(x)=\mu_x+(1-p)\cdot\tau_x$ in the estimation of the Qini curve and AUQ. For simulated data, it is possible to apply this kind of adjustment and so we did it in the simulation settings of our computational experiment. For real-world data, we calculated an estimate $\hat{\Phi}(x)=p\cdot\hat{\mu}_0(x)+(1-p)\cdot\hat{\mu}_{1}(x)$, with $\hat{\mu}_0(x),\hat{\mu}_1(x)$ being outcome models for the untreated respectively treated, trained on data that was not used for the uplift model evaluation. By applying $y-\hat{\Phi}(x)$ as the outcome in the estimation of the Qini curve and AUQ, we also achieve unbiased variance reduction.\citep{bokelmann2023improving}     

\subsection{Variance estimation and calculation of confidence intervals}
To assess the precision of ATE and Qini curve estimates, we needed to calculate the variance (at each decile for the case of the Qini curve). On the simulated data this was quite simple: We just repeated the whole estimation procedure (including the sampling of the data) 1,000 times and then calculated the empirical variance. This was possible, because for simulated data, we could guarantee that the 1,000 data sets were statistically independent. In contrast, for the real-world data it was not possible to sample 1,000 independent (i.e. separate) sub-sets of the whole data set. Hence, the empirical variance between the 1,000 repetitions would underestimate the real variance. For this reason, we applied a slightly different variance estimation strategy than for the simulated data: We applied the variance estimator
\begin{align*}
    \hat{Var}[\hat{ATE}]=\frac{\hat{Var}[y|w=1]}{N_w}+\frac{\hat{Var}[y|w=0]}{N_{\bar{w}}}
\end{align*} in each of the 1,000 repetitions to estimate the variance of the ATE estimator. We then took the average of these 1,000 repetitions to get a precise estimate. Similarly, we estimated the Qini curve variance at each decile, where we applied the same procedure only for $\hat{ATE}_t$ ($t=\frac{j}{10}$ for $j=1,...,10$) instead of $\hat{ATE}$.   

The assessment of uplift model training also required some statistical considerations. As uplift model building is a process involving much variance, the results of one and the same uplift modeling approach vary a lot, depending on the randomly chosen training data set (see right plot in figure \ref{fig:treat_var}). To perform meaningful comparisons between different uplift modeling procedures, this variance needs to be taken into account. For this reason, we decided to calculate confidence intervals for the expected percentage gain in AUQ
\begin{align*}
    E\left[\frac{AUQ(\hat{\tau}_{HS}(x))}{AUQ(\hat{\tau}_R(x))}\cdot 100\%\right],
\end{align*} of an HS-sampled data based uplift model $\hat{\tau}_{HS}(x)$ compared to a completely randomly sampled data-based uplift model $\hat{\tau}_R(x)$. To this end, we repeated the model building procedure (including sampling of training data) 1,000 times. In this way, we had 1,000 observations of the form $\frac{AUQ(\hat{\tau}_{HS}(x))_i}{AUQ(\hat{\tau}_R(x))_i}\cdot 100\%$ ($i=1,...,1000$). The average of these observations was used as an estimator for the expected percentage gain in AUQ. Based on the 1,000 observations, we created confidence intervals for this expected value by assuming asymptotic normality of the average.

\bibliographystyle{unsrtnat}
\bibliography{template}

\end{document}